\documentclass [twocolumn]{article}
\usepackage{amsmath}
\usepackage{amsfonts}
\usepackage{amssymb}
\usepackage{graphicx}
\usepackage{float}
\usepackage{color}

\begin{document}
\title{Controlled Secret Sharing Protocol using a Quantum Cloning Circuit}
\author{Satyabrata Adhikari $^1$ \thanks {satya@iitj.ac.in}, Sovik Roy $^{2, 3}$,
Shantanav Chakraborty $^1$,\\
V. Jagadish $^4$, M. K. Haris $^4$ , Atul Kumar $^1$ \thanks{atulk@iitj.ac.in}
\\ $^1$ Indian Institute of Technology Jodhpur-11, Rajasthan, India \\
$^2$ Techno India, Salt Lake City, Kolkata - 91, India \\ $^3$ S.
N. Bose National Centre for Basic Sciences, Salt Lake,
Kolkata - 98, India \\
$^4$ Indian Institute of Science Education and Research,
Thiruvanathapuram - 16, Kerala, India }

\maketitle

\begin{abstract}
We demonstrate the possibility of controlling the success probability of a secret sharing protocol  using a quantum cloning circuit. The cloning circuit is used to clone the qubits containing the encoded information and {\em en route} to the intended receipients. The success probability of the protocol depends on the cloning parameters used to clone the qubits. We also establish a relation between the concurrence  of initially prepared state, entanglement of the mixed state received by the receivers after cloning scheme and the cloning parameters of cloning machine. 

\end{abstract}

\section{Introduction}
Quantum entanglement \cite{einstein} is purely a quantum mechanical phenomena which has no classical analogue. Apart from being central to the foundational aspects of quantum physics, it has also been used as a resource in communication protocols to perform tasks such as quantum computing \cite{bennett1}, quantum teleportation \cite{bennett2, Zeil97}, quantum cryptography \cite{gisin}, quantum secret sharing \cite{hillery, Cleve99} etc. which are impossible to achieve using classical resources.
In quantum secret sharing, quantum information encoded in a qubit is split among several parties such that only one of them is able to recover the information exactly, provided all the other parties agree to cooperate \cite{hillery}. Quantum secret sharing protocol was carried out using a bipartite pure entangled state in \cite{karlsson} and using tripartite pure entangled states in \cite{bandyopadhyay, bagherinezhad, lance, gordon, zheng}. In \cite{li}, a semi-quantum secret sharing protocol was proposed using maximally entangled GHZ state that was secured against eavesdropping. Quantum secret sharing has also been realized experimentally in \cite{tittel, schmid, schmid1, bogdanski}. In real experimental set ups, the entangled resource shared by the users would be a mixed entangled state due to noise or eavesdropping. We address the question of exploiting the mechanism of noise introduced in the system for gaining advantage in certain situations. \par
In order to explain the results obtained in this article, we first review the original quantum secret sharing protocol- Suppose a spy named Charlie is working under two commanders, Alice and Bob. Charlie's job is to communicate a secret information to both the commanders. However, he suspects that one of the commanders may be dishonest without knowing who among Alice and Bob might it be. Thus, he decides to send the secret information in such a way that one commander cannot collect the information without the help of other. \par
In this article, we consider a situation where there are two secret agents- one honest and one dishonest. The dishonest agent has some information that he/she wants to communicate to two of his/her friends using secret sharing protocol. However, the honest secret agent wants to stop him/her from communicating the information by acting as an eavesdropper and controlling the success of the secret sharing protocol. For this, we assume that Charlie is a dishonest agent who wants to communicate a secret information to his fellow friends Alice and Bob using secret sharing without the knowledge of the agency. Cliff being a loyal and honest agent comes to know about it and wants to stop Charlie from communicating the information. We show that Cliff can in fact stop Charlie from communicating the secret information using a quantum cloning circuit (acting as an eavesdropper) and controlling  the success probability of the protocol. \par

We study secret sharing protocol with the above mentioned problem in focus where the noise is introduced in the system through eavesdropping using a quantum cloning circuit. Our study reveals some interesting facts regarding the relation between the cloning parameters and success probability of the protocol. We show that the bipartite state (sent by Charlie with the encoded classical information) received by Alice and Bob will be an entangled resource iff the concurrence of the initially prepared pure state surpasses a certain threshold value. We further establish that Cliff can always control the success probability of the protocol. Interestingly for a specific quantum cloning scheme, Cliff would succeed in stopping the protocol with certainty using appropriately chosen cloning parameters.  \par
The structure of this article is as follows. In section-II, we study the effect of a quantum cloning circuit on a pure state in $k\times k$-dimensional Hilbert space. For two qubit systems, we find that the mixed state shared between two distant partners remains entangled if the concurrence of the initially prepared pure entangled state exceeds a certain threshold value. In section-III, we discuss the advantages of using a quantum cloning circuit under the scenario discussed in section I. Finally we conclude in section-IV.

\section{Quantum cloning circuit: Creation of mixed states in the Secret Sharing Protocol}
In this section, we study the entanglement properties of a shared state (communicated by Charlie to Alice and Bob) to be used as a resource for secret sharing protocol. For this, we first assume that Charlie prepares and encodes the classical information in a maximally entangled pure state in his laboratory and sends one qubit each to Alice and Bob through two separate channels. The initially prepared pure state in $k\otimes k$-dimensional Hilbert space evolves into a mixed state as the individual qubits  en route (to the users) are cloned by Cliff using a quantum cloning circuit. \vskip 0.1cm
Let us now discuss the evolution of initially prepared pure state into a mixed state. Any bipartite pure state $|\psi\rangle^{in}$ in a $k\otimes k$-dimensional system can be written in the Schmidt polar form as
\begin{eqnarray}
|\psi\rangle^{in}=\sum_{i=1}^{k}\sqrt{\lambda_{i}}|i\rangle_{1}\otimes|i\rangle_{2}
\label{state}
\end{eqnarray}
where $\lambda_{i} \geq 0,~i=1,2,........k$ are the Schmidt coefficients and satisfy the condition $\sum_{i=1}^{k}\lambda_{i}=1$. Once Charlie prepares the bipartite entangled state, he sends particle 1 to Alice and particle 2 to Bob through insecure channels where Cliff attempts to clone the particles $1$ and $2$, respectively. The operation of the quantum cloning circuit that Cliff applies to the individual particles can be described as \cite{buzek}
\begin{equation}
\begin{split}
|i\rangle_{a}|0\rangle_{E}|M\rangle_{x}\rightarrow
c|i\rangle_{a}|i\rangle_{E}|X_{i}\rangle_{x}+\\ d\sum_{j\neq
i}^{k}(|i\rangle_{a}|j\rangle_{E}+|j\rangle_{a}|i\rangle_{E})|X_{j}\rangle_{x}
\end{split}
\label{transformation}
\end{equation}
where $|0\rangle_{E}$ denotes the initial state of the environment and $|M\rangle_{x}$ and $|X_{i}\rangle_{x} (i=1,2,....k)$ denote the ancilla states.
The ancilla state vectors $|X_{i}\rangle_{x} (i=1,2,....k)$ form an orthonormal basis of the ancilla Hilbert space. The unitarity of the transformation
(\ref{transformation}) gives the following relation between the parameters $c$ and $d$
\begin{eqnarray}
c^{2}+2(k-1)d^{2}=1 \label{unitary}
\end{eqnarray}
Using (\ref{transformation}), we presume that Cliff applies
symmetric quantum cloning circuit to both the qubits given in
(\ref{transformation}). The state $|\psi\rangle^{in}$ transforms
under (\ref{transformation}) as
\begin{eqnarray}
\lefteqn{|\psi\rangle^{in}\rightarrow|\psi\rangle^{out}=} & &
\nonumber \\ & &
c^2\sum_{i=1}^{k}\sqrt{\lambda_{i}}\left[|i,i\rangle_{13}\otimes|i,i\rangle_{24}|X_{i}\rangle|X_{i}\rangle\right] \nonumber \\
&+& cd\sum_{i\neq
j}^{k}\sqrt{\lambda_{i}}|i,i\rangle_{13}\otimes{}\nonumber(|i,j\rangle_{24}+|j,i\rangle_{24})|X_{i}\rangle|X_{j}\rangle \nonumber \\
&+& cd\sum_{i\neq j}^{k}\sqrt{\lambda_{i}}(|i,j\rangle_{13}+|j,i\rangle_{13})\otimes|i,i\rangle_{24} |X_{j}\rangle |X_{i}\rangle {}\nonumber \\
&+& d^{2}\sum_{i=1}^{k}\sqrt{\lambda_{i}}[\sum_{i\neq j}^{k}(|i,j\rangle_{13}+ |j,i\rangle_{13})\otimes \sum_{i\neq l}(|i,l\rangle_{24} \nonumber \\
&+& |l,i\rangle_{24})|X_{j}\rangle|X_{l}\rangle].
\label{output}
\end{eqnarray}
where $|\rangle_{3}$ and $|\rangle_{4}$ denote the qubit of the environment.\vskip 0.1cm
After tracing out the ancilla qubits, the four qubit state would be described by the density operator $\rho_{1324}$. Moreover, as the sent qubit 1 (2) interacts with its corresponding cloned qubit 3 (4), the
state described by the density operator $\rho_{13}$ $(\rho_{24})$ can be designated as local output(s) and would be given by
\begin{equation}
\begin{split}
\rho_{13}^{local}=\rho_{24}^{local}=c^{2}\sum_{i=1}^{k}\lambda_{i}|i,i\rangle\langle i,i| +\\d^{2}\sum_{i\neq j}^{k}\lambda_{i}(|i,j\rangle+ |j,i\rangle)(\langle i,j|+\langle j,i|).
\end{split}
\label{local}
\end{equation}
Since the state described by the density operator $\rho_{14}$ $(\rho_{23})$ is formed between the original qubit 1 (2) and the cloned qubit 4 (3) which are located at two distant places, the state can be regarded as a non-local state such that
\begin{eqnarray}
\lefteqn{\rho_{14}^{non-local} = \rho_{23}^{non-local}= } & & \nonumber \\ & &
P\sum_{i=1}^{k}\lambda_{i}|i,i\rangle\langle
i,i|+ Q\sum_{i\neq j}^{k}\sqrt{\lambda_{i}\lambda_{j}} |i,i\rangle\langle j,j| \nonumber \\ &+&
{} R\sum_{i\neq j}^{k}\lambda_{i}(|i,j\rangle\langle i,j| +|j,i\rangle\langle j,i|) \nonumber \\ &+& S\sum_{l,j\neq i}\lambda_{i}|j,l\rangle\langle j,l|
\label{nonlocal}
\end{eqnarray}
where $P=(c^{2}+(k-1)d^{2})^{2}$, $Q=d^{2}(4c^{2}+4cd(k-2)+(k-2)d^{2})$, $R=d^{2}(c^{2}+(k-1)d^{2})$, $S=d^{4}$.
Once Cliff intercepts and clones the particles, he resends the particles 1(2) and 4(3) to Alice and Bob who then share a
joint mixed state described by the density operator $\rho_{14}$ $(\rho_{23})$. It is important to mention here that Cliff may
also send particles 1 (3) and 2 (4), respectively to Alice and Bob. However, without any loss of generality we assume that for Alice and Bob to share an entangled state, Cliff sends particles 1 and 4 or 3 and 2 to Alice and Bob, respectively. This is obvious by the way local and non-local density operators are defined through equations (\ref{local}) and (\ref{nonlocal}). We also demonstrate the viability of above argument by studying the entanglement
properties of density operators defined by (\ref{local}) and (\ref{nonlocal}). For this, we re-express the local and nonlocal outputs
for a $2 \otimes 2$ system in computational basis as
\begin{eqnarray}
\rho_{13}^{local}=\rho_{24}^{local}= \left(%
\begin{array}{cccc}
  c^{2}\lambda_{1} & 0 & 0 & 0 \\
  0 & d^{2} & d^{2} & 0  \\
  0 & d^{2} & d^{2} & 0 \\
  0 & 0 & 0 & c^{2}\lambda_{2}\\
  \end{array}%
\right)
\end{eqnarray}
\begin{eqnarray}
\lefteqn{\rho_{14}^{non-local}= \rho_{23}^{non-local}= } & & \nonumber \\ &
\left(%
\begin{array}{cccc}
  P\lambda_{1}+S\lambda_{2} & 0 & 0 & Q\sqrt{\lambda_{1}\lambda_{2}} \\
  0 & R & 0 & 0  \\
  0 & 0 & R & 0 \\
  Q\sqrt{\lambda_{1}\lambda_{2}} & 0 & 0 & P\lambda_{2}+S\lambda_{1}\\
  \end{array}%
\right)
\end{eqnarray}
where $P=(c^2+d^2)^2, Q=4c^2d^2, R=c^2d^2+d^4, S=d^4$.
As Cliff clones the qubits en route to Alice and Bob using the cloning circuit, the shared mixed state (\ref{nonlocal}) may or may not be entangled.
Using concurrence \cite{hill} and optimal witness operators \cite{bertlmann} for two-qubit systems, we find that the shared state (\ref{nonlocal})
would be entangled if there exists a critical value of concurrence which measures the initial entanglement present in the two qubit pure system
i.e. if the concurrence of initially prepared state is less than this critical value then the shared state is separable.
For this, we use an optimal witness operator $W_{1}^{(2)}$ for a two qubit system given by \cite{bertlmann}
\begin{eqnarray}
W_{1}^{(2)}=\frac{1}{2\sqrt{3}}(I-\vartheta) \label{bwitness},
\end{eqnarray}
where $\vartheta$ can be expressed in terms of the pauli matrices $\sigma_{x},\sigma_{y}~\textrm{and}~\sigma_{z}$ as
\begin{eqnarray}
\vartheta =
\sigma_{x}\otimes\sigma_{x}-\sigma_{y}\otimes\sigma_{y}+\sigma_{z}\otimes\sigma_{z}\label{paulimat}
\end{eqnarray}
In matrix form, $W_{1}^{(2)}$ can be re-expressed as
\begin{eqnarray}
W_{1}^{(2)}= \left(%
\begin{array}{cccc}
  0 & 0 & 0 & \frac{-1}{\sqrt{3}} \\
  0 & \frac{1}{\sqrt{3}} & 0 & 0  \\
  0 & 0 & \frac{1}{\sqrt{3}} & 0 \\
 \frac{-1}{\sqrt{3}} & 0 & 0 & 0\\
  \end{array}%
\right)
\end{eqnarray}
Therefore,
\begin{eqnarray}
\lefteqn{Tr(W_{1}^{(2)}\rho_{14})=Tr(W_{1}^{(2)}\rho_{23})=} & & \nonumber \\ & & (\frac{-2}{\sqrt{3}})(Q\sqrt{\lambda_{1}\lambda_{2}}-R)\label{trace}
\end{eqnarray}
The non-local output $\rho_{14}^{non-local}=\rho_{23}^{non-local}$ would be entangled iff $Tr(W_{1}^{(2)}\rho_{14}) < 0$, hence
\begin{eqnarray}
\begin{split}
Q\sqrt{\lambda_{1}\lambda_{2}}-R>0\Rightarrow
2\sqrt{\lambda_{1}\lambda_{2}}=C(|\psi\rangle^{in})\\>C^{cr}(|\psi\rangle^{in})=\frac{1+c^2}{4c^2},~~~\frac{1}{\sqrt{3}}<c\leq1
\end{split}
\label{cond}
\end{eqnarray}
It is clear that the critical value of concurrence depends on the
cloning parameter c. Also, the function of cloning parameter c in
(\ref{cond}) is a decreasing function and therefore the critical
value of concurrence decreases as c increases. Hence, the lower
value of concurrence (of the initially prepared entangled state)
would ensure that the non-local shared state is entangled if the
quantum cloning circuit parameter c tends towards unity.
Similarly, the local shared state described by the density matrix
$\rho_{13}^{local}=\rho_{24}^{local}$ is separable because
$Tr(W_{1}^{(2)}\rho_{13}^{local})=Tr(W_{1}^{(2)}\rho_{24}^{local})=\frac{1}{3\sqrt{3}}>0$.
Similar results would be obtained if one uses another optimal
witness operator \cite{sanpera} expressed as
\begin{eqnarray}
W_{2}^{(2)}=\frac{1}{2}(I-\vartheta) \label{bwitness1}
\end{eqnarray}
where $\vartheta$ is given by (\ref{paulimat}).
\vskip 0.2cm
\textbf{Observation:} If Charlie initially prepares a maximally entangled state, i.e. when $\lambda_{1}=\lambda_{2}=\frac{1}{2}$,
then for a specific value of parameter $c=\sqrt{2/3}$, the shared state between Alice and Bob takes the form of a maximally entangled
mixed state represented by
\begin{eqnarray}
\rho_{23}^{non-local}=\rho_{14}^{non-local}= \left(%
\begin{array}{cccc}
  \frac{13}{36} & 0 & 0 & \frac{4}{18} \\
  0 & \frac{5}{36} & 0 & 0  \\
  0 & 0 & \frac{5}{36} & 0 \\
  \frac{4}{18} & 0 & 0 & \frac{13}{36}\\
  \end{array}%
\right)\\=\frac{4}{9}|\Phi^{+}\rangle\langle\Phi^{+}|+\frac{5}{36}I_{4}
\label{werner}
\end{eqnarray}
where $|\Phi^{+}\rangle=\frac{1}{\sqrt{2}}(|00\rangle+|11\rangle)$. Thus if the maximally entangled pure state sent through
insecure channels is cloned by Cliff using transformations defined in (\ref{transformation}), then there exists a value of the
quantum cloning circuit parameter c for which the maximally entangled pure state transforms to a maximally entangled mixed state
that belongs to the family of Werner states \cite{werner}.
\section{Application of two-qubit bipartite mixed state in the quantum secret sharing protocol}
We now proceed to demonstrate the application of secret sharing protocol so that Cliff can stop Charlie from communicating secret message encoded in the two-qubit mixed entangled state (\ref{nonlocal}) shared between Alice and Bob. Our protocol can be described in following steps: \vskip 0.5cm

\textbf{Step-I: Maximally entangled pure state prepared by Charlie}\vskip 0.3cm

In order to split the information between Alice and Bob, Charlie prepares a two
qubit maximally entangled pure state either in  $|\phi^{+}\rangle=\frac{1}{\sqrt{2}}(|00\rangle+|11\rangle)$ or in  $|\phi^{-}\rangle=\frac{1}{\sqrt{2}}(|00\rangle-|11\rangle)$ form. Whether to prepare $|\phi^{+}\rangle$ or $|\phi^{-}\rangle$ state is decided by tossing a coin i.e. if ``head" appears then Charlie prepares $|\phi^{+}\rangle$ and if ``tail" appears then Charlie prepares $|\phi^{-}\rangle$. One can designate ``head" as ``0" and ``tail" as ``1". In this way, Charlie encodes one bit of information into the prepared state. Once the information is encoded, Charlie sends the qubits to Alice and Bob. Cliff, however, intercepts and clones these qubits using a quantum cloning circuit. For both the qubits, Cliff applies the same symmetric quantum cloning circuit described by (\ref{transformation}). The protocol proceeds with Cliff resending any one of the two qubits to Alice and Bob provided the state is entangled. Without any loss of generality we assume that Alice and Bob share a mixed state that can be described either by the density operator
\begin{eqnarray}
\rho_{AB}^{+} &=&
\frac{P+S}{2}(|00\rangle\langle00|+|11\rangle\langle11|) \nonumber \\ &+& \frac{Q}{2}(|00\rangle\langle11|+|11\rangle\langle00|) \nonumber \\ &+& R(|01\rangle\langle01|
+|10\rangle\langle10|)
\label{shared state1}
\end{eqnarray}
or by the density operator
\begin{eqnarray}
\rho_{AB}^{-} &=&
\frac{P+S}{2}(|00\rangle\langle00|+|11\rangle\langle11|) \nonumber \\ &-& \frac{Q}{2}(|00\rangle\langle11|+|11\rangle\langle00|) \nonumber \\ &+& R(|01\rangle\langle01|
+|10\rangle\langle10|)
\label{shared state2}
\end{eqnarray}
where $P=(c^{2}+d^{2})^{2}$, $Q=4c^{2}d^{2}$, $R=d^{2}(c^{2}+d^{2})$, $S=d^{4}$ and $c^2+2d^2=1$.
\vskip 0.5cm

\textbf{Step-II: Single qubit measurements performed by Alice}\vskip 0.3cm

Alice performs measurement on her qubit in the Hadamard basis
$B_{H}=\{\frac{|0\rangle+|1\rangle}{\sqrt{2}},\frac{|0\rangle-|1\rangle}{\sqrt{2}}\}$. The single qubit state received by Bob would depend on the measurement outcome of Alice's qubit. For example, \vskip
0.1cm




(i) If the shared state between Alice and Bob is $\rho_{AB}^{+}$ and Alice's measurement outcome is $\frac{|0\rangle+|1\rangle}{\sqrt{2}}$, then
\begin{eqnarray}
\rho_{B}^{+0} &=& Tr_{1}\left[\left(\frac{|0\rangle+|1\rangle}{\sqrt{2}}\frac{\langle0|+\langle1|}{\sqrt{2}}\otimes I_{2}\right)\rho_{AB}^{+}\right. \nonumber \\ & & \left.\left(\frac{|0\rangle+|1\rangle}{\sqrt{2}}\frac{\langle0|+\langle1|}{\sqrt{2}}\otimes I_{2}\right)\right]{}\nonumber \\ &=& \frac{1}{2}\left[I_{2}+Q(|0\rangle\langle1|+|1\rangle\langle0|)\right].
\label{outcome1}
\end{eqnarray}
(ii) If the shared state between Alice and Bob is $\rho_{AB}^{+}$ and Alice's measurement outcome is $\frac{|0\rangle-|1\rangle}{\sqrt{2}}$, then
\begin{eqnarray}
\rho_{B}^{+1} &=& Tr_{1}\left[\left(\frac{|0\rangle-|1\rangle}{\sqrt{2}}\frac{\langle0|-\langle1|}{\sqrt{2}}\otimes I_{2}\right)\rho_{AB}^{+}\right. \nonumber \\ & & \left.\left(\frac{|0\rangle-|1\rangle}{\sqrt{2}}\frac{\langle0|-\langle1|}{\sqrt{2}}\otimes I_{2}\right)\right]{}\nonumber \\ &=& \frac{1}{2}\left[I_{2}-Q(|0\rangle\langle1|+|1\rangle\langle0|)\right].
\label{outcome2}
\end{eqnarray}
(iii) If the shared state between Alice and Bob is $\rho_{AB}^{-}$ and Alice's measurement outcome is $\frac{|0\rangle+|1\rangle}{\sqrt{2}}$, then
\begin{eqnarray}
\rho_{B}^{-0} &=& Tr_{1}\left[\left(\frac{|0\rangle+|1\rangle}{\sqrt{2}}\frac{\langle0|+\langle1|}{\sqrt{2}}\otimes I_{2}\right)\rho_{AB}^{-}\right. \nonumber \\ & & \left.\left(\frac{|0\rangle+|1\rangle}{\sqrt{2}}\frac{\langle0|+\langle1|}{\sqrt{2}}\otimes I_{2}\right)\right]{}\nonumber \\ &=& \frac{1}{2}\left[I_{2}-Q(|0\rangle\langle1|+|1\rangle\langle0|)\right].
\label{outcome3}
\end{eqnarray}
(iv) If the shared state between Alice and Bob is $\rho_{AB}^{-}$ and Alice's measurement outcome is $\frac{|0\rangle-|1\rangle}{\sqrt{2}}$, then
\begin{eqnarray}
\rho_{B}^{-1} &=& Tr_{1}\left[\left(\frac{|0\rangle-|1\rangle}{\sqrt{2}}\frac{\langle0|-\langle1|}{\sqrt{2}}\otimes I_{2}\right)\rho_{AB}^{-}\right. \nonumber \\ & & \left.\left(\frac{|0\rangle-|1\rangle}{\sqrt{2}}\frac{\langle0|-\langle1|}{\sqrt{2}}\otimes I_{2}\right)\right]{}\nonumber \\ &=& \frac{1}{2}\left[I_{2}+Q(|0\rangle\langle1|+|1\rangle\langle0|)\right].
\label{outcome4}
\end{eqnarray}
where $I_{2}$ denotes the identity operator in $2\times2$-dimensional Hilbert space. Similarly, one can find the state obtained by Alice, if Bob chose to perform measurement on his qubit. The equations (\ref{outcome1}) - (\ref{outcome4}) clearly explain that it is neither possible for Alice nor for Bob alone to decode Charlie's encoded information. They would only be able to decode Charlie's information if they both agree to collaborate with each other. If they agree to collaborate, then our protocol proceeds further to step-III.
\vskip 0.3cm

\textbf{Step-III: Alice declares her measurement outcome}\vskip 0.3cm

Once Alice and Bob agree to cooperate with each other, Alice sends her measurement outcome to Bob. \vskip 0.1cm

(i) If the measurement outcome is $\frac{|0\rangle+|1\rangle}{\sqrt{2}}$ then she sends Bob a
classical bit $``0"$ and \vskip 0.1cm

(ii) If the measurement outcome is $\frac{|0\rangle-|1\rangle}{\sqrt{2}}$ then she sends classical bit
$``1"$ to Bob. \vskip 0.3cm

\textbf{Step-IV: Positive operator valued measurement (POVM) performed by Bob}\vskip 0.3cm

Here, we introduce the positive operators to unambiguously discriminate between Bob's mixed states $\rho_{B}^{+0}$ and $\rho_{B}^{-0}$ (or $\rho_{B}^{+1}$ and $\rho_{B}^{-1}$) corresponding to Alice's measurement outcome $\left|+\right\rangle$ (or $\left|-\right\rangle$). For this, we define three-element positive operator valued measurement (POVM), denoted by $E_{1}$, $E_{2}$ and $E_{3}$ such that
\begin{eqnarray}
\begin{array}{l}
{E_1} = \left( {\begin{array}{*{20}{c}}
{\frac{Q}{2}}&1\\
0&{\frac{Q}{2}}
\end{array}} \right)\\
{E_2} = \left( {\begin{array}{*{20}{c}}
{\frac{Q}{2}}&{ - 1}\\
0&{\frac{Q}{2}}
\end{array}} \right)\\
{E_3} = I - {P_1} - {P_2}
\label{povm}
\end{array}\end{eqnarray}
where $Q \in [0, 1/2]$.
If Bob receives the classical bit $``0"$ then Alice's measurement outcome would be $\left|+\right\rangle$ and correspondingly Bob will receive either $\rho_{B}^{+0}$ or $\rho_{B}^{-0}$. Using the operators defined in (\ref{povm}), Bob can successfully discriminate between the two states $\rho_{B}^{+0}$ and $\rho_{B}^{-0}$ as
\begin{eqnarray}
\begin{array}{l}
Tr({E_1}\rho _B^{ - 0}) = Tr({E_2}\rho _B^{ + 0}) = 0\\
Tr({E_1}\rho _B^{ + 0}) = Tr({E_2}\rho _B^{ - 0}) = Q
\end{array}
\end{eqnarray}
Similarly, if Bob receives the classical bit $``1"$ then also he can discriminate the single qubit states $\rho_{B}^{+1}$ and $\rho_{B}^{-1}$ using POVM operators defined in (\ref{povm}). \vskip 0.1cm
Thus, the success probability of the protocol can be expressed as
\begin{equation}
\begin{split}
P_{suc}=\frac{1}{2}Tr[\rho^{+0}E_1]+\frac{1}{2}Tr[\rho^{-0}E_2].
\end{split}
\end{equation}
The above equation can be re-expressed as
\begin{equation}
P_{suc}= Q = 4c^{2}d^{2},
\end{equation}
where $Q\in[0,\frac{1}{2}]$. Clearly, the success probability depends on the cloning circuit parameters $c$ and $d$ such that $P_{suc}\leq 1/2$.
 Therefore, the success probability of the protocol can be controlled by Cliff and would lie between $Q\in[0,\frac{1}{2}]$.
 For universal quantum cloning machine where the cloning parameters $c$ and $d$ assume the values  $\frac{2}{\sqrt{3}}$ and $\frac{1}{\sqrt{6}}$,
 respectively $P_{suc}$ = $0.45$ which is very close to the maximum success probability $(1/2)$ that can be achieved using this protocol.
 However, if Cliff wants to stop Charlie from communicating the secret information to Alice and Bob, he will use Wootters-Zurek
 \cite{wootters}
 cloning machine where the cloning parameters $c$ and $d$ take values  $1$ and $0$, respectively and hence the success probability
 of the protocol would be zero. In this way, Cliff can control the success probability of the protocol and can stop Charlie from
 leaking the secret information.
\section{Conclusion}
Our study provides a different insight to the secret sharing protocol in a scenario where an honest secret agent wants to stop a dishonest agent from leaking certain information using secret sharing protocol. We have shown that by using a cloning circuit the honest agent can control the success probability of secret sharing protocol and in a specific case can even stop the dishonest agent with certainty from communicating the secret information. It would be interesting to see how Cliff can control the success probability of the protocol by using a more generic cloning transformation in comparison to (\ref{transformation}). Another question of particular interest would be to analyze the protocol for transmission of a quantum information instead of a classical one.

\vskip 0.3cm.

\end{document}